\documentclass[sigconf]{acmart}
\AtBeginDocument{%
  }

\setcopyright{acmlicensed}
\copyrightyear{2018}
\acmYear{2018}
\acmDOI{XXXXXXX.XXXXXXX}
\acmConference[Conference acronym 'XX]{Make sure to enter the correct
  conference title from your rights confirmation email}{June 03--05,
  2018}{Woodstock, NY}
\acmISBN{978-1-4503-XXXX-X/2018/06}

\usepackage{xcolor}
\usepackage{pifont}
\usepackage{enumitem}
\usepackage{url}
\usepackage[most]{tcolorbox}

\definecolor{codebg}{RGB}{248,248,248}
\definecolor{keyword}{RGB}{0,0,255}
\definecolor{stringcolor}{RGB}{163,21,21}
\definecolor{keycolor}{RGB}{0,102,204}

\lstdefinestyle{jsonstyle}{
  basicstyle=\ttfamily\tiny,
  breaklines=true,
  numbers=none,
  frame=none,
  escapeinside={(*@}{@*)},  
  backgroundcolor=\color{white!10},
  literate=
    *{0}{{{\color{blue}0}}}{1}
     {1}{{{\color{blue}1}}}{1}
     {2}{{{\color{blue}2}}}{1}
     {3}{{{\color{blue}3}}}{1}
     {4}{{{\color{blue}4}}}{1}
     {5}{{{\color{blue}5}}}{1}
     {6}{{{\color{blue}6}}}{1}
     {7}{{{\color{blue}7}}}{1}
     {8}{{{\color{blue}8}}}{1}
     {9}{{{\color{blue}9}}}{1}
     {:}{{{\color{red}:}}}{1}
     {,}{{{\color{red},}}}{1}
     {"}{{{\color{orange}"}}}{1}
}
\newcommand{\opencoderrank}{OpenCoderRank}
\lstdefinestyle{textstyle}{
    backgroundcolor=\color{white},
    basicstyle=\ttfamily\tiny,
    numbers=none,
    frame=none,
    breaklines=true,
    breakindent=0pt,
    breakatwhitespace=true,
    showstringspaces=false,
    captionpos=b,
    escapeinside={(*@}{@*)},  
    language={}
}
\copyrightyear{2026}
\acmYear{2026}
\setcopyright{cc}
\setcctype{by}
\acmConference[SynthIR@SIGIR 2026]{Proceedings of The First Workshop on Synthetic Content in Information Retrieval Ecosystems (SynthIR@SIGIR 2026)}{July 24, 2026}{Melbourne, VIC, Australia}

\acmBooktitle{Proceedings of The First Workshop on Synthetic Content in Information Retrieval Ecosystems (SynthIR@SIGIR 2026), July 24, 2026, Melbourne, VIC, Australia}
\begin{document}

\title{\opencoderrank: Personalized Technical Assessments with Generative AI}

\author{Hridoy Sankar Dutta}

\orcid{1234-5678-9012}
\authornotemark[1]
\affiliation{%
  \institution{Deakin University}
  \city{GIFT City}
  \country{India}
}
\email{hridoy.dutta@deakin.edu.au}

\author{Sana Ansari}
\affiliation{%
  \institution{S.P. Jain Institute of Management and Research}
  \city{Mumbai}
  \country{India}
}
\email{sana.ansari@spjimr.org}

\author{Swati Kumari}
\affiliation{%
  \institution{Deakin University}
  \city{GIFT City}
  \country{India}
}
\email{swati.kumari@deakin.edu.au}

\author{Shounak Ravi Bhalerao}
\affiliation{%
  \institution{Deakin University}
  \city{GIFT City}
  \country{India}
}
\email{shounakbhalerao777@gmail.com}

\renewcommand{\shortauthors}{Dutta et al.}

\begin{abstract}
Organizations and educational institutions use time-bound assessment tasks to evaluate coding and problem-solving skills. These assessments measure not only the correctness of the solutions, but also their efficiency. Problem setters (educator/interviewer) are responsible for crafting these challenges, carefully balancing difficulty and relevance to create meaningful evaluation experiences. Conversely, problem solvers (student/interviewee) apply critical and logical thinking to arrive at correct solutions. In the era of Large Language Models (LLMs), LLMs assist problem setters in generating diverse and challenging questions, but they can undermine assessment integrity for problem solvers by providing easy access to solutions.  We introduce \opencoderrank\, a lightweight, self‑hosted platform that emulates real‑world timed technical assessments in resource‑constrained environments. \opencoderrank\ is intentionally model‑agnostic: it facilitates the creation, deployment and automatic grading of problems while offering fine‑grained control over time limits, input–output pairs and execution constraints. \opencoderrank\  is evaluated using two methods: 1. \textit{BERTScore}, 2. LLM evaluation. Our findings indicate that \opencoderrank\ connects problem setters and solvers by supporting time-constrained preparation and self-hosted, customizable assessments in resource-constrained settings.
\end{abstract}




\maketitle

\section{Introduction}
With the rise in Large Language Models (LLMs), they have become valuable tools for problem setters, helping them create a wide variety of complex and well-structured questions more efficiently \cite{desmond2025evalassist,verga2024replacing}. However, problem solvers increasingly utilize LLMs, thereby creating challenges for problem setters in maintaining the fairness of technical assessments through readily quick code generation and creating an uneven playing field \cite{bhushan2025detecting,fleckenstein2024teachers,koike2024outfox}.  The technical assessments often include time-bound assessments that test both correctness and performance against tests. For example, in academic setting, to help students  measure their problem-solving ability, universities often tie up with cost-intensive third-party platforms or outsource assessment services to external vendors. Without a controlled environment, students can easily open multiple tabs or access external tools during assessments, undermining the validity of test results. These issues cannot be mitigated effectively unless the assessment platform enforces restrictions to ensure legitimate assessment practices such as disabling copy-paste, enforcing full screen mode, etc.

\begin{figure*}[!t]
    \centering
    \includegraphics[width=0.9\textwidth]{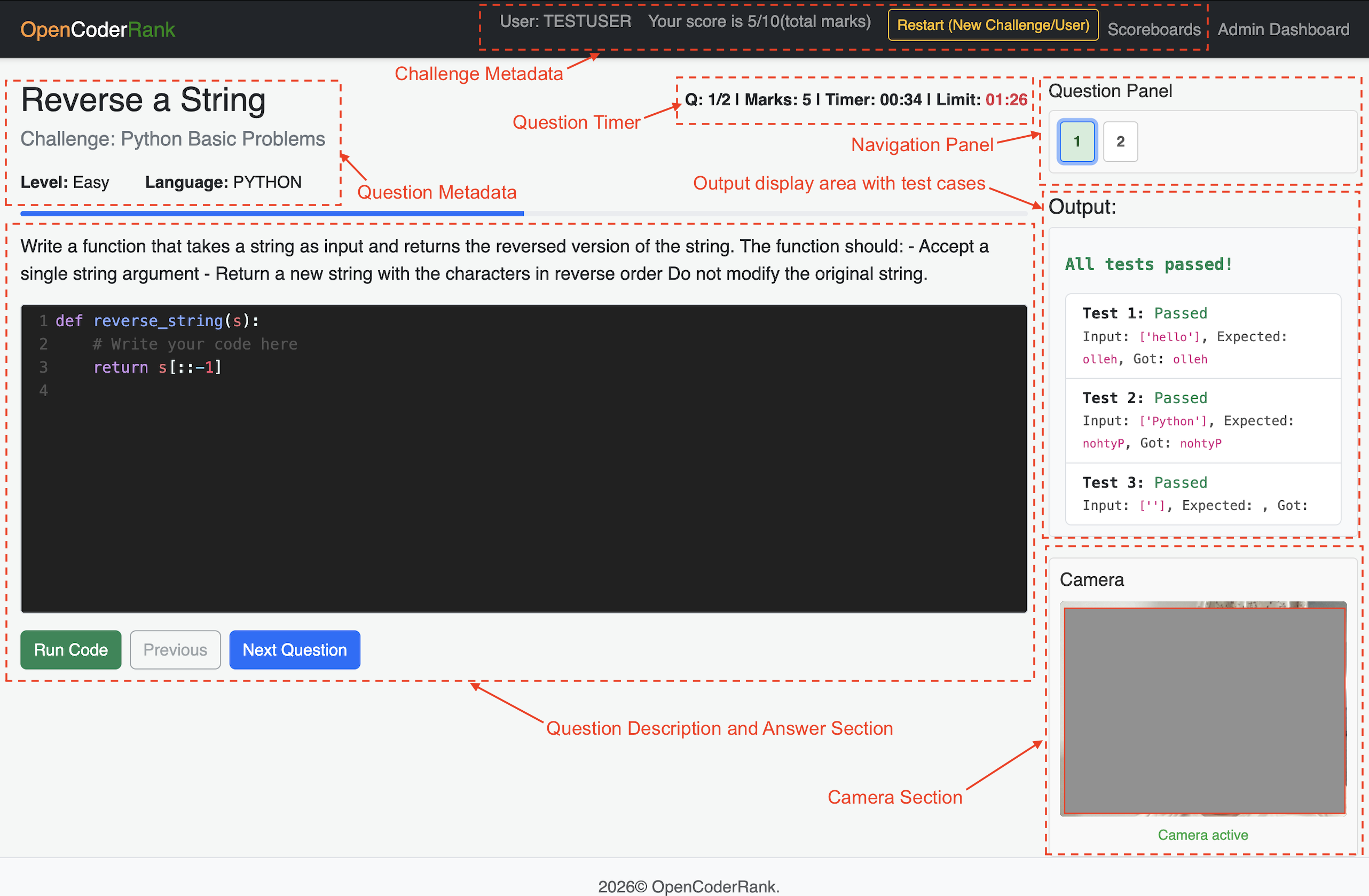}
    \caption{Overview of the \opencoderrank\ interface with key features labeled.}
    \label{fig:overview}
\end{figure*}

To address these challenges, we present \opencoderrank\ (Source Code\footnote{\url{https://anonymous.4open.science/r/OpenCoderRank-D13E/}}, Video tutorial\footnote{\url{https://anonymous.4open.science/r/OpenCoderRank-D13E/OpenCoderRank_video.mp4}} and Tool Prototype\footnote{\url{https://opencoderrank.onrender.com/}}), a platform designed to host assessment tasks that simulate industry and academic style challenges to reduce the gap between problem setters and problem solvers. Figure \ref{fig:overview} shows an overview of \opencoderrank\ with key features labeled. \opencoderrank\ aims to answer two research questions.
\begin{itemize}[leftmargin=*, noitemsep,topsep=0pt]
    \item \textbf{RQ1:} How can a self-hosted assessment platform improve pedagogy in a resource-constrained environment?
    \item \textbf{RQ2:} How can a self-hosted assessment platform maximize its effectiveness in the era of LLMs?
\end{itemize}
We use the terms \textit{interviewers, educators} and \textit{recruiters} interchangeably to refer to problem setters on the \opencoderrank\ platform, and the terms \textit{interviewees, students} and \textit{participants} to refer to problem solvers.

\section{Related Work}
Educators and interviewers around the world are increasingly using LLMs to create innovative methods to teach programming and develop the technical skills of students \cite{chu2025llm, zhang2024simulating, chu2025uniedu}. For example, \cite{puadurean2024automated} introduced BugSpotter, which uses LLMs to auto-generate buggy Python code exercises that help students develop proficiency in bug localization by designing failing test cases.  \cite{liffiton2023codehelp} developed CodeHelp, an LLM-powered tool that provides on-demand programming assistance without directly revealing solutions. \cite{kannam2025code} designed live oral coding interviews to evaluate student proficiency based on take-home work. Besides technical assessments, LLMs have also been explored for other technical educational purposes such as reducing educator workload \cite{del2024evaluating,jordan2024need,phung2023generative}, teaching students debugging strategies for programming \cite{kafai2019rethinking,whalley2021novice}, etc. We provide a detailed comparison of \opencoderrank\ with existing tools in the literature, as summarized in Table \ref{tab:comparison}.

\noindent \textbf{\underline{Our study:}} While the above mentioned tools focus on enhancing learning and developing skills, they do not fully address critical challenges in technical assessments in the current scenario. Motivation for industry is different, recruiters primarily aim to test whether students possess fundamental problem-solving skills and can think critically under pressure. To evaluate this, companies often rely on targeted programming tasks designed to test whether students arrive at the correct output efficiently. This outcome-focused evaluation contrasts with many existing educational tools that prioritize guiding the student through the learning process rather than simulating real-world test conditions. Hence, there is a gap between what students practice in academic settings and what they are expected to demonstrate during industry hiring. 
\begin{table}[!htbp]
\centering
\caption{Comparison of \opencoderrank\ with existing tools in the literature}
\label{tab:comparison}
\begin{tabular}{p{4cm}|c|c|c|c|c}
\hline
\textbf{Feature / Tool} & 
\cite{puadurean2024automated} &
\cite{liffiton2023codehelp} &
\cite{kannam2025code} &
\cite{desmond2025evalassist} &
\textbf{Ours} \\ 
\hline

LLM Question Generation & 
{\color{green}\ding{52}} & 
{\color{red}\ding{55}} & 
{\color{red}\ding{55}} & 
{\color{green}\ding{52}} & 
{\color{green}\ding{52}} \\ 
\hline

Automated Judging / Scoring & 
{\color{red}\ding{55}} & 
{\color{red}\ding{55}} & 
{\color{red}\ding{55}} & 
{\color{green}\ding{52}} & 
{\color{green}\ding{52}} \\ 
\hline

Self-Hosted / Local Deployment & 
{\color{red}\ding{55}} & 
{\color{red}\ding{55}} & 
{\color{red}\ding{55}} & 
{\color{red}\ding{55}} & 
{\color{green}\ding{52}} \\ 
\hline

Integrity Controls (Fullscreen, Copy-Paste Lock) & 
{\color{red}\ding{55}} & 
{\color{red}\ding{55}} & 
{\color{red}\ding{55}} & 
{\color{red}\ding{55}} & 
{\color{green}\ding{52}} \\ 
\hline

Randomized Question Sequencing & 
{\color{red}\ding{55}} & 
{\color{red}\ding{55}} & 
{\color{red}\ding{55}} & 
{\color{red}\ding{55}} & 
{\color{green}\ding{52}} \\ 
\hline

Programming/MCQ modes & 
{\color{red}\ding{55}} & 
{\color{green}\ding{52}} & 
{\color{green}\ding{52}} & 
{\color{red}\ding{55}} & 
{\color{green}\ding{52}} \\ 
\hline
Camera Feature  & 
{\color{red}\ding{55}} & 
{\color{red}\ding{55}} & 
{\color{red}\ding{55}} & 
{\color{red}\ding{55}} & 
{\color{green}\ding{52}} \\ 
\hline

Resource-Constrained Friendly & 
{\color{red}\ding{55}} & 
{\color{red}\ding{55}} & 
{\color{red}\ding{55}} & 
{\color{red}\ding{55}} & 
{\color{green}\ding{52}} \\ 
\hline
\end{tabular}

\end{table}
\opencoderrank\ bridges the gap between problem-setters and problem-solvers by combining the flexibility of learning platforms with the rigor of industry-style assessments. It ensures academic integrity, rapid setup and deployment and a framework for creating customized challenges.

\section{\opencoderrank\ Overview}
\begin{figure}
    \centering
    \includegraphics[width=\linewidth]{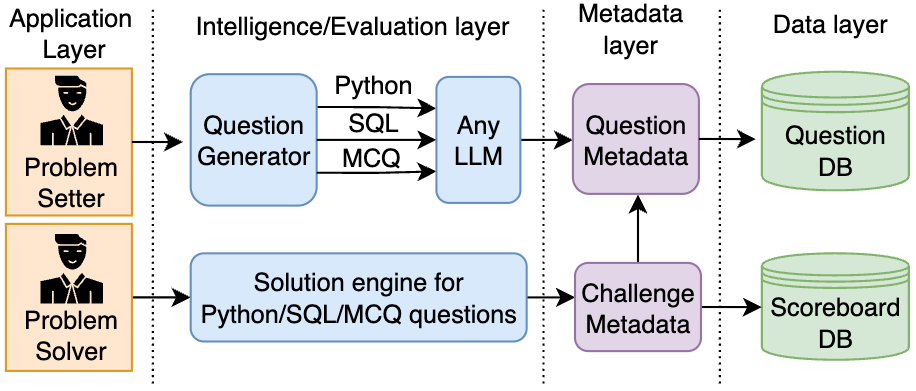}
    \caption{High-level pipeline of \opencoderrank}
    \label{fig:pipeline}
\end{figure}

In this section, we will discuss key technical details of \opencoderrank. The overall pipeline of \opencoderrank\ is presented in Figure \ref{fig:pipeline} spanning application, intelligence/evaluation, metadata and data layers. A problem setter creates questions via a generator that can leverage any LLM for Python, SQL or MCQ tasks, while a solution engine evaluates submissions from problem solvers. Generated questions and challenges are stored with relevant metadata in dedicated databases. \opencoderrank\ is built with the Python Flask\footnote{\url{https://flask.palletsprojects.com/en/stable/}} framework and uses SQLite\footnote{\url{https://sqlite.org/index.html}} as the database.

\subsection{Key Features}
\opencoderrank\ offers a comprehensive set of features for both problem setters and problem solvers. Problem setters can create, edit and manage problems and problem solvers can engage effectively with the problems. Each problem includes
\begin{itemize}[leftmargin=*]
    \item \textbf{Challenge details:} Related \textit{Challenge ID} for the problem.
    \item \textbf{Problem details:} \textit{Title}, \textit{description} and \textit{level (easy/medium/hard)} of the problem.
    \item \textbf{Language:}  \textit{mcq} for MCQ questions, \textit{python/sql} for coding questions.
    \item \textbf{Input/output specifications:} \textit{Valid/invalid options} and \textit{Correct answer} for MCQ questions, \textit{starter code} and \textit{test cases} for coding-based assessments, \textit{time limit} and \textit{points} for both MCQ and coding-based assessments.
        \item \textbf{Remarks:} Any \textit{remarks/hints} related to the problem.
\end{itemize}
\opencoderrank\ also supports an ``Camera'' feature that allows problem setters to guide problem solvers during assignment solving.

\subsection{Question Generator and Judging} 
\begin{figure}[!htbp]
    \centering
    \includegraphics[width=\linewidth]{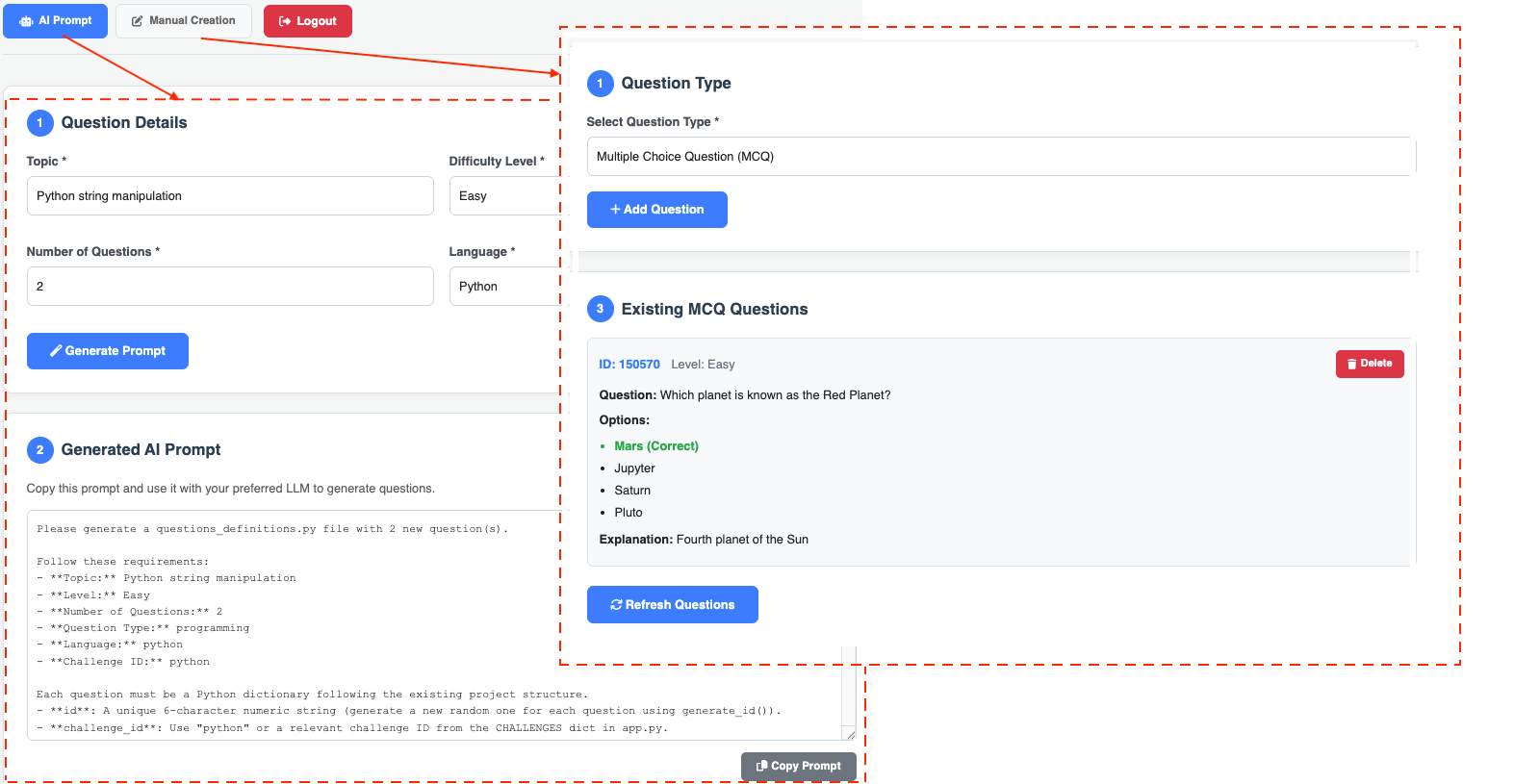}
    \caption{Question Generator with the prompt to generate questions on any LLM (a longer and more detailed version of the question generator prompt is present in the GitHub repository).}
    \label{fig:question-generator}
\end{figure}

Inspired by Chain-of-Thought reasoning \cite{wei2022chain}, \opencoderrank\ encourages more structured and transparent reasoning in the creation and evaluation of technical assessments. \opencoderrank\ is developed primarily as an assessment platform. Problem setters can design their own set of questions or generate one question to create new, increasingly difficult questions using the AI prompt. \opencoderrank\  supports seamless, web-based question generation that is fully embedded into the workflow, removing the need for manual copy-and-paste of prompts as shown in Figure \ref{fig:question-generator}. For educators from non-technical backgrounds, we additionally provide a manual question creation interface specifically for MCQ questions. \opencoderrank\ supports automated judging of MCQ questions and supports two programming languages - Python and SQL.
\subsection{Rapid Setup and Deployment} One of the key strengths of \opencoderrank\ is its simplicity and speed of deployment. The platform is built for quick setup and can be self-hosted in just two straightforward steps. The other alternatives are to host it on cloud-based platforms\footnote{\url{https://www.netlify.com/}} for broader access or share it through tunneling tools\footnote{\url{https://pinggy.io/}} to run it with minimal overhead. This makes it especially valuable for organizers who need a fast, flexible and reliable solution for events with minimal configuration. \opencoderrank\ includes Docker support that allows users to containerize the entire application and deploy it consistently across different systems. This makes it especially valuable for problem setters who need a fast, flexible and reliable solution for events with minimal configuration.

\subsection{Evaluation}
\opencoderrank\ is model-agnostic and does not depend on any specific LLM. We therefore evaluate the quality of assessment content hosted on the platform, rather than the performance of any particular generation model, similar to the approach followed by Molavi et al. \cite{molavi2025llm}. We assess the quality of \opencoderrank-generated technical assessment questions using a combination of embedding-based semantic similarity and model-based automatic evaluation. Specifically, we employ BERTScore \cite{zhang2020bertscore} to measure semantic alignment between generated questions and expert-authored problems from LeetCode and HackerRank. BERTScore leverages contextual embeddings from pretrained language models to enable robust comparison beyond surface-level lexical overlap. We also adopt an LLM-as-a-Judge evaluation framework following recent work on automated model-based evaluation \cite{desmond2025evalassist,verga2024replacing}, in which a state-of-the-art large language model is prompted as an expert computer science educator to assess each generated question along four dimensions: \textit{clarity} (readability and unambiguity), \textit{correctness} (technical validity and logical consistency), \textit{pedagogical usefulness} (instructional and assessment value) and \textit{difficulty alignment} (appropriateness relative to the intended difficulty level).

We generate the questions using GPT-4o and evaluate them using the Gemini 3 Flash model\footnote{\url{https://docs.cloud.google.com/vertex-ai/generative-ai/docs/models/gemini/3-flash}} within our automated generation and evaluation pipeline. From an evaluation of 100 technical assessment questions sourced from LeetCode and HackerRank, we observe that \opencoderrank\ produces high-quality and pedagogically meaningful content that closely aligns with expert-authored problems, as summarized in Table~\ref{tab:evaluation}. We use a different model for evaluation to reduce bias, since models tend to favor outputs similar to their own style. We observe that semantic similarity analysis using BERTScore yields an F1 score of 0.79, indicating substantial alignment with reference problems.
\begin{table}[htbp]
\centering
\caption{Evaluation of \opencoderrank-generated assessment questions compared to expert-authored problems from LeetCode and HackerRank.}
\label{tab:evaluation}
\begin{tabular}{lcc}
\hline
\textbf{Evaluation Aspect} &
\textbf{Score} &
\textbf{Mechanism} \\
\hline
Semantic Similarity (BERTScore) & 0.79/1 & vs. expert-authored \\
Question Clarity & 4.98/5 & LLM-as-a-Judge \\
Question Correctness & 4.93/5 & LLM-as-a-Judge \\
Pedagogical Usefulness & 4.92/5 & LLM-as-a-Judge \\
Difficulty Alignment  & 4.39/5 & LLM-as-a-Judge \\
\hline
\end{tabular}
\end{table}
To further assess question quality, as mentioned above, we employ an LLM-as-a-Judge framework. This evaluation confirms the strong quality of the generated questions, with high average scores for clarity (4.98/5) and correctness (4.93/5), reflecting well-structured and technically sound problem statements. Pedagogical usefulness achieves a score of 4.92/5, demonstrating that the questions are well suited for instructional and assessment purposes. In addition, the difficulty alignment score (4.39/5) indicates that the generated problems generally match their intended difficulty levels.

\section{Answering the research questions}
\subsection{RQ1: Pedagogical Implications} \opencoderrank\ is designed to enhance learning outcomes across a variety of teaching and evaluation contexts. Below are several pedagogical use cases where \opencoderrank\ can be effectively used:
\begin{itemize}[leftmargin=*]
    \item \textbf{Coding Competitions}: 
    \opencoderrank\ provides a unique environment for hosting intra- or inter-college coding competitions. Various coding clubs within academic institutions can also deploy it locally for quick peer learning exercises. With features like fast setup and real-time evaluation, the platform enables organizers to run contests even in resource-constrained environments with zero setup costs.
    \item \textbf{Classroom Integration}:
    Although \opencoderrank\ is primarily focused on coding and programming exercises, the platform also supports MCQs, making it suitable for quick assessments or pre/post-lecture quizzes. Educators can incorporate \opencoderrank\ into regular classroom teaching as part of time-constrained lab sessions or live quizzes. 
    \item \textbf{Academic Lab/Startup Hiring}: 
    The easy setup of \opencoderrank\ allows academic research labs and startups from various domains to conduct technical assessments for hiring assistants or interns. Custom challenges aligned with specific focus areas (e.g., MCQs on domain knowledge or relevant programming tasks) can be used to evaluate candidate abilities prior to subsequent rounds of the interview process.
    \item \textbf{One-to-One/One-Vs-One Challenges}: 
    The platform supports personalized one-to-one coding challenges, which are especially useful in mentoring sessions or during office hours. Educators can assign individualized tasks and provide real-time feedback to create a focused learning experience.
\end{itemize}

\subsection{RQ2: Ensuring Integrity in the LLM Era:} The rise of LLMs poses new challenges for maintaining integrity in technical assessments. To encounter this, \opencoderrank\ implements several strategies to mitigate this risk: 
\begin{itemize}[leftmargin=*]
\item \textbf{Full-screen mode enforcement}: \opencoderrank\ is designed to run only in full-screen mode. When participants click the ``Start Test'' button, a prompt informs them that the assessment must be taken in full-screen mode to minimize distractions. If a participant exits or disables full-screen mode five times, the test is automatically submitted.
    \item \textbf{Disabling copy-paste}: \opencoderrank\ promotes legitimate assessment practices, such as writing code without relying on external tools by preventing direct copying from LLMs or external sources.
    \item \textbf{Random sequencing of questions}: Questions are presented in a randomized order making it difficult to rely on neighboring screens during assessments. 
\end{itemize}
\section{Conclusion and Future Work}
This paper presents \opencoderrank, an LLM-powered platform designed to overcome the limitations of traditional technical assessments. \opencoderrank\ serves as a pedagogical bridge between problem setters and problem solvers, enabling students to practice under timed conditions and tackle unfamiliar problems, while empowering problem setters to self-host and customize assessments. By ensuring integrity for technical assessments, \opencoderrank\  adapts to modern evaluation challenges while preserving fairness in the era of LLMs.

\opencoderrank\ represents a step forward in personalized education, offering customized creation and evaluation for designing and evaluating technical assessments and by setting the stage for future innovations in adaptive learning methodologies. In future work, we plan to incorporate more advanced user-profiling signals such as skill trajectories, problem-solving behavior patterns and difficulty adaptation curves to provide deeper insight and more personalized assessment experiences. The range of assessment types supported by \opencoderrank\ will continue to expand as well. We also aim to address key technical challenges to safeguard integrity in the LLM era that uphold assessment integrity in \opencoderrank\ by focusing on robust anti-cheating strategies and reliable verification mechanisms.

\bibliographystyle{acm}
\bibliography{sample-base}

@inproceedings{kannam2025code,
  title={Code Interviews: Design and Evaluation of a More Authentic Assessment for Introductory Programming Assignments},
  author={Kannam, Suhas and Yang, Yuri and Dharm, Aarya and Lin, Kevin},
  booktitle={Proceedings of the 56th ACM Technical Symposium on Computer Science Education V. 1},
  pages={554--560},
  year={2025}
}

@article{puadurean2024automated,
  title={Automated Generation of Code Debugging Exercises},
  author={P{\u{a}}durean, Victor-Alexandru and Denny, Paul and Singla, Adish},
  journal={arXiv e-prints},
  pages={arXiv--2411},
  year={2024}
}

@inproceedings{liffiton2023codehelp,
  title={Codehelp: Using large language models with guardrails for scalable support in programming classes},
  author={Liffiton, Mark and Sheese, Brad E and Savelka, Jaromir and Denny, Paul},
  booktitle={Proceedings of the 23rd Koli Calling International Conference on Computing Education Research},
  pages={1--11},
  year={2023}
}

@article{chu2025llm,
  title={Llm agents for education: Advances and applications},
  author={Chu, Zhendong and Wang, Shen and Xie, Jian and Zhu, Tinghui and Yan, Yibo and Ye, Jinheng and Zhong, Aoxiao and Hu, Xuming and Liang, Jing and Yu, Philip S and others},
  journal={arXiv preprint arXiv:2503.11733},
  year={2025}
}

@article{zhang2024simulating,
  title={Simulating classroom education with llm-empowered agents},
  author={Zhang, Zheyuan and Zhang-Li, Daniel and Yu, Jifan and Gong, Linlu and Zhou, Jinchang and Hao, Zhanxin and Jiang, Jianxiao and Cao, Jie and Liu, Huiqin and Liu, Zhiyuan and others},
  journal={arXiv preprint arXiv:2406.19226},
  year={2024}
}

@article{chu2025uniedu,
  title={UniEDU: A Unified Language and Vision Assistant for Education Applications},
  author={Chu, Zhendong and Xie, Jian and Wang, Shen and Wang, Zichao and Wen, Qingsong},
  journal={arXiv preprint arXiv:2503.20701},
  year={2025}
}

@inproceedings{desmond2025evalassist,
  title={Evalassist: Llm-as-a-judge simplified},
  author={Desmond, Michael and Ashktorab, Zahra and Geyer, Werner and Daly, Elizabeth M and Cooper, Martin Santillan and Pan, Qian and Nair, Rahul and Wagner, Nico and Pedapati, Tejaswini},
  booktitle={Proceedings of the AAAI Conference on Artificial Intelligence},
  volume={39},
  number={28},
  pages={29637--29639},
  year={2025}
}

@article{verga2024replacing,
  title={Replacing judges with juries: Evaluating llm generations with a panel of diverse models},
  author={Verga, Pat and Hofstatter, Sebastian and Althammer, Sophia and Su, Yixuan and Piktus, Aleksandra and Arkhangorodsky, Arkady and Xu, Minjie and White, Naomi and Lewis, Patrick},
  journal={arXiv preprint arXiv:2404.18796},
  year={2024}
}

@inproceedings{koike2024outfox,
  title={Outfox: Llm-generated essay detection through in-context learning with adversarially generated examples},
  author={Koike, Ryuto and Kaneko, Masahiro and Okazaki, Naoaki},
  booktitle={Proceedings of the AAAI Conference on Artificial Intelligence},
  volume={38},
  number={19},
  pages={21258--21266},
  year={2024}
}

@article{fleckenstein2024teachers,
  title={Do teachers spot AI? Evaluating the detectability of AI-generated texts among student essays},
  author={Fleckenstein, Johanna and Meyer, Jennifer and Jansen, Thorben and Keller, Stefan D and K{\"o}ller, Olaf and M{\"o}ller, Jens},
  journal={Computers and Education: Artificial Intelligence},
  volume={6},
  pages={100209},
  year={2024},
  publisher={Elsevier}
}

@article{bhushan2025detecting,
  title={Detecting LLM-Generated Short Answers and Effects on Learner Performance},
  author={Bhushan, Shambhavi and Thomas, Danielle R and Borchers, Conrad and Raghuvanshi, Isha and Abboud, Ralph and Gatz, Erin and Gupta, Shivang and Koedinger, Kenneth},
  journal={arXiv preprint arXiv:2506.17196},
  year={2025}
}

@inproceedings{del2024evaluating,
  title={Evaluating automatically generated contextualised programming exercises},
  author={del Carpio Gutierrez, Andre and Denny, Paul and Luxton-Reilly, Andrew},
  booktitle={Proceedings of the 55th ACM Technical Symposium on Computer Science Education V. 1},
  pages={289--295},
  year={2024}
}

@inproceedings{jordan2024need,
  title={Need a programming exercise generated in your native language? chatgpt's got your back: Automatic generation of non-english programming exercises using openai gpt-3.5},
  author={Jordan, Mollie and Ly, Kevin and Soosai Raj, Adalbert Gerald},
  booktitle={Proceedings of the 55th ACM Technical Symposium on Computer Science Education V. 1},
  pages={618--624},
  year={2024}
}

@inproceedings{phung2023generative,
  title={Generative AI for programming education: benchmarking ChatGPT, GPT-4, and human tutors},
  author={Phung, Tung and P{\u{a}}durean, Victor-Alexandru and Cambronero, Jos{\'e} and Gulwani, Sumit and Kohn, Tobias and Majumdar, Rupak and Singla, Adish and Soares, Gustavo},
  booktitle={Proceedings of the 2023 ACM Conference on International Computing Education Research-Volume 2},
  pages={41--42},
  year={2023}
}

@inproceedings{kafai2019rethinking,
  title={Rethinking debugging as productive failure for CS education},
  author={Kafai, Yasmin B and DeLiema, David and Fields, Deborah A and Lewandowski, Gary and Lewis, Colleen},
  booktitle={Proceedings of the 50th ACM technical symposium on computer science education},
  pages={169--170},
  year={2019}
}

@inproceedings{whalley2021novice,
  title={Novice reflections on debugging},
  author={Whalley, Jacqueline and Settle, Amber and Luxton-Reilly, Andrew},
  booktitle={Proceedings of the 52nd ACM technical symposium on computer science education},
  pages={73--79},
  year={2021}
}

@article{wei2022chain,
  title={Chain-of-thought prompting elicits reasoning in large language models},
  author={Wei, Jason and Wang, Xuezhi and Schuurmans, Dale and Bosma, Maarten and Xia, Fei and Chi, Ed and Le, Quoc V and Zhou, Denny and others},
  journal={Advances in neural information processing systems},
  volume={35},
  pages={24824--24837},
  year={2022}
}

@inproceedings{zhang2020bertscore,
  title={BERTScore: Evaluating Text Generation with BERT},
  author={Zhang, Tianyi and Kishore, Varsha and Wu, Felix and Weinberger, Kilian Q. and Artzi, Yoav},
  booktitle={International Conference on Learning Representations (ICLR)},
  year={2020}
}

@inproceedings{molavi2025llm,
  title={LLM-Driven Personalized Answer Generation and Evaluation},
  author={Molavi, Mohammadreza and Tavakoli, Mohammadreza and Moein, Mohammad and Faraji, Abdolali and Kismih{\'o}k, G{\'a}bor},
  booktitle={International Conference on Artificial Intelligence in Education},
  pages={187--194},
  year={2025},
  organization={Springer}
}

\end{document}